\documentclass[twocolumn, ceqn, prb, showpacs, showkeys, superscriptaddress]{revtex4-1}

\usepackage{amssymb,amsmath,xcolor,graphicx,xspace,colortbl,rotating} 
\usepackage{amssymb,amsmath,xcolor,graphicx,xspace,colortbl,rotating}  
\usepackage{tabulary}  
\usepackage{textcomp}  
\usepackage{varioref}  
\usepackage{units}
\DeclareGraphicsExtensions{.eps,.ps,.png,.pdf,.jpg,.jpeg}

\begin{document}
\title{Vortex Variable Range Hopping in a Conventional Superconducting Film }
\author{Ilana M. Percher }
\affiliation{School of Physics and Astronomy, University of Minnesota, Minneapolis, MN 55455, USA }
\author{Irina Volotsenko}
\affiliation{Department of Physics, Bar Ilan University, Ramat Gan, ISRAEL}
\author{Aviad Frydman }
\affiliation{Department of Physics, Bar Ilan University, Ramat Gan, ISRAEL}
\author{ Boris I. Shklovskii }
\affiliation{School of Physics and Astronomy, University of Minnesota, Minneapolis, MN 55455, USA }
\author{ Allen M. Goldman }
\affiliation{School of Physics and Astronomy, University of Minnesota, Minneapolis, MN 55455, USA }
\date{ \today }

\begin{abstract}The behavior of a disordered amorphous thin film of  superconducting Indium Oxide has been studied as a function of temperature and magnetic field applied perpendicular to its plane.  A superconductor-insulator transition has been observed, though the isotherms do not cross at a single point.  The curves of resistance vs. temperature on the putative superconducting side of this transition, where the resistance decreases with decreasing temperature, obey two-dimensional Mott variable-range hopping of vortices over wide ranges of temperature and resistance. To estimate the parameters of hopping, the film is modeled as a granular system and the hopping of vortices is treated in a manner analogous to hopping of charges. The reason the long range interaction between vortices over the range of magnetic fields investigated does not lead to a stronger variation of resistance with temperature than that of two-dimensional Mott variable-range hopping remains unresolved.    
\end{abstract}

\maketitle
\section*{Introduction}
The superconductor-insulator-transition (SIT) of two-dimensional (2D) or quasi-2D films is usually envisioned as a direct zero-temperature quantum phase transition tuned by disorder, magnetic field, or charge density. Because quantum fluctuations associated with a quantum critical point persist at nonzero temperatures, features of a zero-temperature continuous transition can be revealed through measurements of physical properties at accessible temperatures.\cite{LIN}

In the dirty boson model of the SIT a zero-temperature metallic state only exists at the quantum critical point (QCP) with a universal resistance of $h/4 e^{2}$.\cite{FISH} In the case of the field-tuned transition, the magnetic field at the crossing point of the magnetoresistance isotherms is usually taken as the critical field corresponding to the QCP. Here, we report on the resistance vs. temperature of a highly disordered InO\textsubscript {x} film in magnetic fields below and above the critical field of the quantum phase transition. We find below the critical field a temperature dependence consistent with a model involving 2D Mott variable range hopping (VRH)~\cite{MOTT} of vortices, where the vortices move by quantum mechanical tunneling. Above the transition we see a signature of VRH of Cooper pairs.  

The creep and flow of vortices have been subjects of intense study because they bring about nonzero electrical resistance even below superconducting critical temperatures and critical magnetic fields.\cite{AND} Interest in these phenomena surged with the discovery of high temperature oxide superconductors. The motivation for this research was the necessity of controlling pinning so as to assure the zero-resistance required for applications. In the classical picture, vortices are pinned at zero temperature in a glass phase. At finite temperatures they become thermally activated above the glass barriers (de-pinned), resulting in nonzero resistance governed by an Arrhenius form.\cite{BLAT} This behavior is found when the temperature is below but of the order of the temperature scale set by the pinning energy. At lower temperatures, there is the possibility of quantum tunneling of vortices through energy barriers resulting in what might be called quantum flux creep. This subject was first studied by Caldeira and Leggett, who investigated quantum tunneling of vortices in superconducting quantum interference devices,\cite{CLD} by Glazman and Fogel', who considered vortex tunneling in very thin films,\cite{GLAZ} and by Mitin,\cite{MIT} who considered vortex tunneling in bulk superconductors. There is limited experimental literature on quantum tunneling of vortices in films of either conventional~\cite{LIU,EPH} or high temperature superconductors.\cite{SEF,KOR} Experimentally, vortex tunneling in superconducting films would appear as variable range hopping.

Quantum variable-range hopping of bundles of vortices in disordered 2D superconductors at low temperatures and high magnetic fields was considered by Fisher, Tokayasu, and Young (FTY), who predicted a decreasing resistivity with decreasing temperature as $\rho  \propto \exp  [ -(T_{1}/T)^{p}]$, where $p$ is in the range from 2/3 to 4/5, and $T_{1}$ is a characteristic temperature.\cite{FTY}  Shklovskii suggested an alternative theory of single vortex variable range hopping which leads to a temperature dependence similar to FTY.\cite{BOR} These predictions were supported by the work of Sefrioui \textit{et al.} on deoxygenated YBa\textsubscript {2}Cu\textsubscript {3}O\textsubscript {6.4} thin films at high fields and low current densities.\cite{SEF} Auerbach, Arovas, and Gosh (AAG) calculated the tunneling rate of a single vortex between two pinning sites and the subsequent resistivity due to flux tunneling at low fields and low temperatures.\cite{AUE} The result was a resistivity for a 2D BCS superconductor following a modified Mott variable range hopping form in 2D,
\begin{equation} R  \propto \exp \left[ -\left( \frac{T_{0}}{T}\right) ^{1/3} \right], \end{equation}where
\begin{equation}T_{0} =\frac{\beta }{k_{B} g (\mu ) a^{2}}\text{,}
\end{equation}and $g (\mu )$ is the density of states of vortices at their chemical potential $\mu $ (see explanation below), $a$ is the vortex localization length, and $\beta  \sim 13$ is a numerical coefficient.\cite{SEBOOK} In contrast with hopping of charge carriers, the hopping of vortices results in a decrease in resistance with decreasing temperature, rather than a increase. This behavior was reported by Koren \textit{et al.} for a YBa\textsubscript {2}Cu\textsubscript {3}O\textsubscript {7- $\delta $} thin film meander line, but over a very limited range of resistances and applied magnetic fields.\cite{KOR} Effects such as these have been observed in conventional superconducting films such as InO\textsubscript {x}, but again over a very limited range of resistances.\cite{BREZ}

\section*{Experimental Methods}
The InO\textsubscript {x} film under study was \unit[55]{nm} thick, grown by electron-beam evaporation of In\textsubscript {2}O\textsubscript {3}. During deposition, an O\textsubscript {2 }partial pressure of \unit[6.7x10\textsuperscript{-5}]{mbar} was maintained in the chamber by bleeding the gas through a needle valve while continuing to pump.\cite{OVAD1} The substrate temperature was kept below about 40\textsuperscript {$\circ$}C so that the films remained amorphous. Measurements were made using an Oxford Instruments Kelvinox 25 dilution refrigerator with the film connected in a van der Pauw configuration. The measuring currents were kept below \unit[15]{nA} for low fields and \unit[5]{nA} at high fields to ensure linear current-voltage characteristics over the whole range of temperatures and magnetic fields.

\begin{figure}[hbtp]\centering 
\includegraphics[width=0.48\textwidth ]{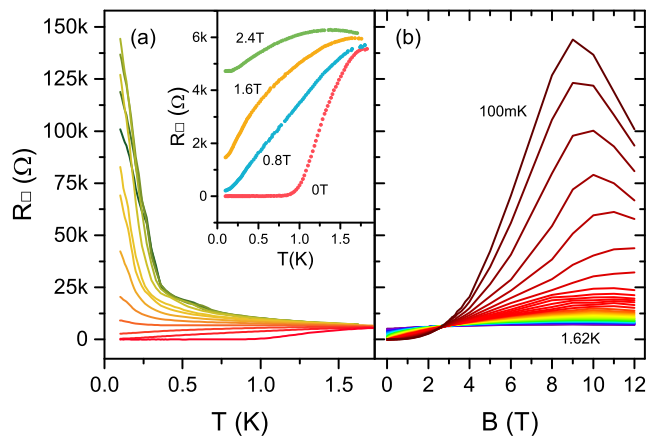}
\caption{(Color online) (a) Temperature dependence of sheet resistance at magnetic fields \unit[0]{T}(bottom) to \unit[12]{T} in increments of \unit[1]{T}. The inset provides a clearer view of the superconducting transition and low-field data. (b) Magnetic field dependence of sheet resistance as a function of temperature from \unit[100]{mK} to \unit[1.62]{K} in increments of \unit[40]{mK}.}\label{RvsT}
\end{figure}

\section*{Results}
Figure~\ref{RvsT}(a) shows curves of $R (T)$ over a range of magnetic fields applied perpendicular to the plane of an InO\textsubscript {x} film. This film exhibited a level of disorder that placed it close to the disorder-tuned superconductor-insulator transition. The inset of the figure shows the very broad resistive transition to the superconducting state in zero magnetic field, indicating a highly disordered film. Figure~\ref{RvsT}(b) shows isotherms of $R (B)$ at various temperatures extracted from the data of Fig.~\ref{RvsT}(a). The isotherms appear to cross at a field $B_c \simeq$ \unit[2.8]{T}, and at a sheet resistance close to that of the quantum resistance for pairs, $h/4 e^{2}$. A standard finite-size scaling analysis was not carried out because a detailed examination of the isotherms revealed that they did not intersect at a single well-defined field. Also for magnetic fields below the critical tuning field, for a region within the the nominally ``superconducting" branch of the data, the resistance did not extrapolate to zero resistance in the $T\rightarrow 0$ limit. The resistance begins to saturate at the lowest temperatures on both sides of the transition. Behavior similar to this on the superconducting side was reported first by Mason and Kapitulnik many years ago, and is referred to as the intermediate metallic regime.\cite{MAS} It has also been reported over the past decade in dc measurements of the magnetic-field-tuned SIT of Ta, InO\textsubscript{x}, exfoliated NbSe\textsubscript{2}, disorder-tuned NbSi, and ionic-liquid gated ZrNCl films.\cite{QIN,Wei,TSEN,COU,SAI} It has been interpreted as evidence of a Bose metal.\cite{DAS1,DAS2,DALID,PHI}  Saturation at lowest temperatures in the insulating regime is likely due to a failure to cool the sample.

\begin{figure}[btp]\centering 
\includegraphics[width=0.48\textwidth ]{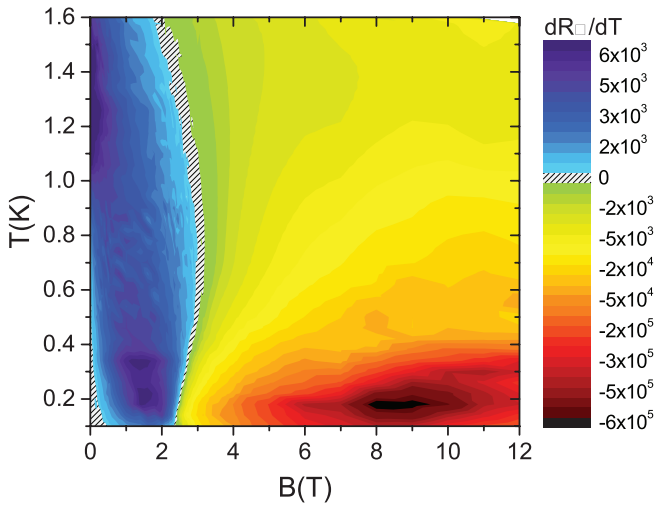}
\caption{(Color online) False-color contour plot showing $d R/d T$ as a function of temperature and magnetic field, which functions as a phase diagram for this sample. The striped region indicates where $d R/d T \approx 0$, which occurs in the lower left due to superconductivity, and in a curved vertical band between \unit[2]{T} and \unit[3]{T}, which corresponds to the transition between superconducting and insulating behavior. To the left of this band, blue and purple (dark) indicate $d R/d T >0$, where behavior is metallic. Yellows and reds indicate $d R/d T <0$, the insulating regime to the right of the band. Within this region, the magnetoresistance peak is responsible for the dark feature at low temperatures.}\label{colorplot}
\end{figure}

One can obtain a clearer picture of the various regimes of this film from a false color plot of $d R/d T$ as a function of temperature and magnetic field, as shown in Fig.~\ref{colorplot}. There is a small region of superconductivity, which shows up as zero slope at temperatures below about \unit[0.5]{K} and at fields below about \unit[0.5]{T}. There is a very wide region of magnetic fields over which there is a positive $d R/d T$. This is separated from negative slope regions found at higher fields by a curved narrow vertical band of zero $d R/d T$. This film exhibits a magnetoresistance peak at low temperatures, which shows up as a region of very large $d R/d T$ at a magnetic field of around \unit[9]{T}.

\begin{figure}[hbtp]\centering 
\centering
\includegraphics[width=0.48\textwidth ]{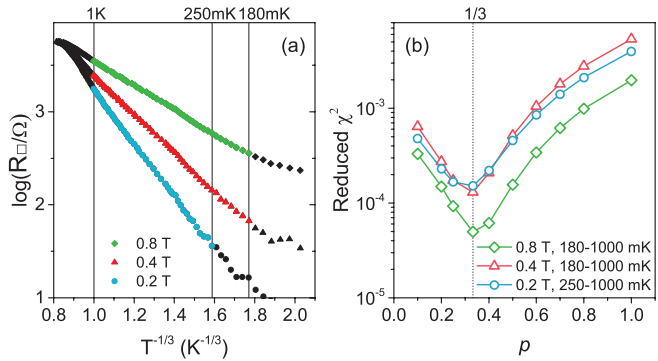}
\caption{(Color Online)(a) The logarithm of sheet resistance plotted as a function of temperature to the power of -1/3, measured at three representative fields.  (b) The value of $\widetilde{\chi}^2$ for linear fits to the data shown in part (a), when plotted as $\log (R)$ vs $T^{-p}$ for different values of $p$.  These fits were performed over the temperature intervals mentioned above and highlighted in part (a).} \label{Fig:fit_figure}
\end{figure}

We then set about to determine whether we could understand the systematic behavior in the regime of positive $d R/d T$ at fields $B < B_c$, by plotting $\log  R$ vs $T^{-p}$, with $p$ taking on various values such as $1 , 2/3 , 1/2, 1/3$ and $1/4$. Then, the fit of the low-field $R$ vs. $T$ data to the form $R \propto \exp  [ -(T_{0}/T)^{p}]$ was evaluated by plotting $\log(R_{\square}/\Omega)$ vs. $T^{-p}$ for different values of $p$, and in each case fitting the data to a line.  The quality of this linear fit was evaluated using $\widetilde{\chi}^2$ (a.k.a. reduced $\chi ^{2}$), the normalized sum of squared deviations between the data and fit function.\cite{TAY}  Fitting was performed on data for the temperature range beginning at $\approx$ \unit[200]{mK}, where the resistance begins to flatten, up to \unit[1]{K}.  These ranges are marked on Fig.~\ref{Fig:fit_figure}(a), which show data measured at several representative low fields for $p=1/3$.  This value of $p$ was found to minimize $\chi^2$ over these temperature intervals, as shown in Fig.~\ref{Fig:fit_figure}(b).

\begin{figure}[tb] \centering
\includegraphics[width=0.48\textwidth ]{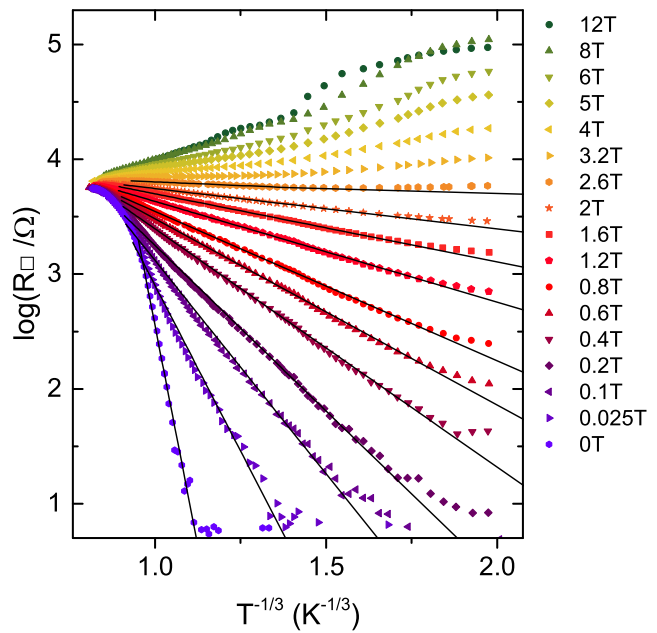}
\caption{(Color online) Logarithm of resistance plotted against $T^{-1/3}$ at various magnetic fields. The straight lines are fits of the data for $B \leq B_c \simeq$ \unit[2.8]{T} to Eq. (1), the functional form for 2D Mott VRH of vortices.} \label{linearfits}
\end{figure}

In Fig.~\ref{linearfits} we show a plot of $\log  R$ vs. $T^{ -1/3}$ for the full range of fields.  Extended regions of resistance vs. temperature at various applied magnetic fields are quantitatively consistent with a 2D Mott variable range hopping law Eq. (1) down to temperatures where the resistance saturates. The saturation resistance found at the lowest temperatures was independent of the measuring current. 

\begin{figure}[bht] \centering
\includegraphics[width=0.32\textwidth]{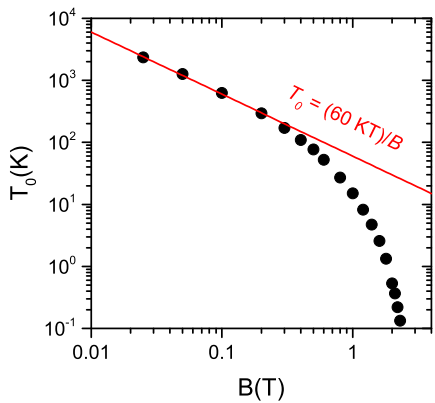}
\caption{(Color online) The characteristic temperature defined in Eqs. (1) and (2), $T_{0}$, as a function of magnetic field.  These points were extracted from fits like those shown in Fig.~\ref{linearfits}.  The straight line shows the $1/B$ dependence of $T_{0}$ at low fields.} \label{T_0vsB}
\end{figure}

The functional dependence of the resistance on temperature is that which was derived for the resistance due to VRH of vortices of a 2D superconductor by AAG.\cite{AUE} A fitting procedure based on Eq. (1) was used to determine $T_{0}$ as a function of $B$. The results are shown in Fig.~\ref{T_0vsB}. Examination of this figure shows that $T_{0}$ initially varies as $C/B$, where $C \simeq$ \unit[60]{TK}, and then falls to zero at a magnetic field corresponding to the critical crossing field of the isotherms. This form for the dependence of $T_{0}$ on magnetic field is not predicted by AAG. Clearly a model of disorder and pinning of vortices in a film is needed to explain the data. 

\begin{figure}[hbtp]\centering 
\centering
\includegraphics[width=0.48\textwidth ]{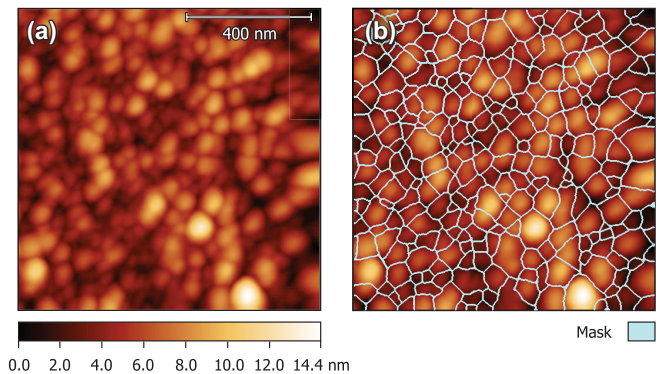}
\caption{(Color Online) (a) Atomic force microscope image of a $\unit[1]{\mu m} \times \unit[1]{\mu m}$ portion of the surface of the InO\textsubscript{x} film under study. The false-color scale indicates the height distribution of the surface of the film, which is on average 55 nm thick. (b) shows the same micrograph superimposed with the grain boundaries used for analysis.}
\label{Fig:AFM}\end{figure}

Atomic force microscope scans of the roughly \unit[55]{nm}-thick film reveal the presence of grain-like surface structures. A Bruker Nanoscope V Multimode 8 was used to image the surface of the film in several different, randomly-chosen locations.  A representative micrograph is shown in Fig. \ref{Fig:AFM}(a).  It shows that the surface is rough and gives the appearance of a spherical granular structure with a spread in grain diameters.

Characteristics of the grains were determined by demarcating them on each of the AFM micrographs, such as is shown in Fig. \ref{Fig:AFM}(b).   Grains were found to have a mean diameter of 51 nm with a standard deviation of 20 nm. 

\section*{Discussion}
Below we treat the film as an idealized random, densely packed two-dimensional array of superconducting spheres with diameters varying around \unit[50]{nm}. The good match between the average lengthscale of the grains and the overall film thickness supports the single layer spherical-grain model used in this paper.  The height data presented above do not suggest that the film is comprised of literal spheres, but are consistent with a granular structure that can be modeled in this way.

The spheres in the model are barely touching each other and are connected via Josephson junctions of very small area. In such a Josephson junction array vortex cores are localized in empty spaces between spheres, which play the role of pinning sites for vortices. Self-energies of vortices in such pinning sites $\epsilon $ vary due to a random distribution of Josephson junction strengths. We assume that pinning sites have the density of states $g (\epsilon )$ with a small characteristic width $\delta \epsilon$ which we address later. We assume that the self-energy of a vortex carrying two flux quanta in one site (double vortex) is larger than this width and it, therefore, does not allow two quantum vortices to occupy a single pinning site. Thus, at low temperatures, single quantum vortices occupy the lowest-energy sites. At fields $B \leqslant $ \unit[0.2]{T}, which are representative of the range of fields where $T_0 \sim 1/B$, the vortex spacing is \unit[140]{nm} or greater, so that no more than 13\% of lattice pinning sites are occupied by vortices. This means that at $T =0$ they occupy only the low-energy tail of the $g (\epsilon )$ of localized pinning sites. The energy of the highest pinning state occupied by a vortex at $T = 0$ plays the role of vortex chemical potential $\mu$. 

Vortex transport under the influence of a superconducting current can then be described as quantum hopping of vortices from occupied to empty pinning sites. At high enough temperatures, vortices hop to nearest neighbor sites. At lower temperatures, vortex transport is constrained to longer-distance hops with energies $\epsilon$ in a narrow band of energies around chemical potential $\mu $. In a homogeneous bulk superconductor a vortex is a massive, almost a classical object with a very small localization length. But in a Josephson junction array where intergrain Josephson energy and the Coulomb charging energy are comparable, the localization length of a vortex $a$ is of the order of the sphere diameter.\cite{IOF} (Of course, in the close vicinity of the SIT, $a$ diverges.) This is because the vortex core can easily tunnel across the weak Josephson contacts formed where neighboring spheres touch, while avoiding the bulk of superconductor. The combination of disorder in vortex energies and relatively large localization length $a$ allows us to think about long-distance vortex hops, and to arrive at the two-dimensional Mott variable range hopping law Eq. (1) with the characteristic temperature given by Eq. (2). Here $g (\mu )$ is the density of states of vortices at their chemical potential $\mu$. 

Let us now show that the observed dependence $T_{0} (B) =C/B$ in small magnetic fields can be interpreted as a result of $g (\mu )$ growing linearly with $B$. Indeed, using a simple model of the low-energy tail of the density of states $g (\epsilon ) =g_{0} \exp [ - (\epsilon_0 - \epsilon) /\delta \epsilon ]$, where $\epsilon_0$ is average energy of the vortex and $\delta \epsilon $ is the dispersion of the vortex self-energy, we find that the two dimensional concentration of vortices
\begin{equation}n_{v} =e B/h c = \smallint _{ -\infty }^{\mu }g (\epsilon ) d \epsilon =g (\mu ) \delta \epsilon .
\end{equation} 
Combining Eqs. (2) and (3) we arrive at the experimental dependence $T_{0} =C/B$, where
\begin{equation} 
C = 2\pi \beta (c\hbar/e^2) \frac{\delta\epsilon}{k_B} \frac{e}{a^2}. 
\end{equation}
In principle, the localization length $a$ depends on $\mu $ and, therefore, on $B$. However, the logarithmic dependence of $\mu $ on $B$ is very weak and is ignored below. Although above we used a simple exponential tail for the density of states, one can easily show that our conclusion that $T_{0} =C/B$ is approximately (with logarithmic accuracy) correct for a Gaussian or other fast-decaying exponential tail.

Let us now estimate the vortex self-energy dispersion $\delta\epsilon$. It is convenient to think about our disordered Josephson junction array as a thin film of thickness $d$ made from a layer of material which has a 3D penetration length $\lambda$.~\cite{Tinkham} The self-energy of a Pearl vortex in such a film is~\cite{PEARL,DGEN}
\begin{equation}
\epsilon_0 = \epsilon_{00}\ln \left (\frac{\lambda_{\perp}}{d}\right),
\end{equation}
where 
\begin{equation}
\epsilon_{00} = \left (\frac{\phi_0}{4\pi}\right)^2 \lambda_{\perp}^{-1}.
\end{equation}
Here $\lambda_{\perp}  = \lambda^2/d$ is the Pearl penetration length of the film with width $d$ in a magnetic field perpendicular to the film.  It truncates the logarithmic divergence of $\epsilon_0$ at large distances, while at small distances the divergence is truncated by the radius of the hole between neighboring spheres, which is of the order of sphere diameter $d$. 
For an ideally periodic Josephson junction~\cite{Tinkham}  $\lambda_{\perp} = c \phi_0^2/(8\pi^2 I_c) $, where $I_c$ is the critical current of a junction, and Eq. (5) is reduced~\cite{Tinkham} to the standard expression $\epsilon_0 = (\pi \hbar I_c/2e)\ln(\lambda^2/d^2)$. 

Let us suggest an interpretation of the origin of the logarithm in Eq. (5), which we use below to estimate $\delta\epsilon$. Here, the vortex energy results from currents in a disc with radius $\lambda_{\perp}$. One can partition the area of this disc into annuli by defining $ \ln(\lambda_{\perp}/d) $ concentric bounding circles.  These have radii $\lambda_{\perp}/2$ , $\lambda_{\perp}$/4, $ \lambda_{\perp}/8$ and so on, until we reach the minimum radius $d$. Currents decay with distance from the center in such a way that each annulus bound by two consecutive circles contributes $\epsilon_{00}$ to the vortex energy. The contribution of the central cell, where the core of the vortex is located, is of the order of $\epsilon_{00}$ as well. Thus, for the the total energy of the vortex we arrive at Eq. (5).

The dispersion in self-energy $\delta\epsilon$ from site to site in a strongly disordered granular film is probably dominated by the dispersion of the critical current $I_c$ of individual Josephson junctions, which are very sensitive to the contact conductance. Below we assume that $\delta I_c/I_c \sim 1$. Thus, the central site alone results in dispersion $\delta\epsilon = \epsilon_{00}$. Fluctuations of $I_c$ averaged over the many Josephson junctions in each large-radius annulus significantly cancel each other. Therefore, the contribution of these annulii to $\delta\epsilon$ are much smaller than that of the central cite and can be neglected. Substituting Eq. (6) for $\delta\epsilon$ into Eq. (4) and using $c\hbar/e^2 = 137$, we arrive at a final result for $C$:
\begin{equation}
C \simeq 5 \times 10^{7} \frac{e^2}{d k_B} \frac{e}{a^2} \left( \frac{d}{\lambda}\right)^{2}.
\end{equation}
Assuming that $a \simeq d=$ \unit[50]{nm} we arrive at the experimental value $C =$ \unit[60]{TK} if the penetration length $\lambda \sim \unit[10]{\mu m}$, which seems reasonable for such a weakly coupled granular superconductor.

So far we have dealt with an analysis of the regime of low magnetic fields (positive $dR/d T$). Let us now switch our attention to the high magnetic field side $(B > B_c)$ of Fig.~\ref{linearfits} (i.e. the insulating regime or negative $d R/d T$). In the insulating regime, one can envision VRH of Cooper pairs of electrons or just single electrons. If one neglects their Coulomb interaction this should lead to conventional Mott VRH with a positive sign in the exponent of Eq. (1). We see that qualitatively such an explanation seems to work. At the highest fields the magnetoresistance changes sign, possibly because of the destruction of Cooper pairs and a crossover to VRH of single electrons.\cite{TIA,LEE}

In the range of fields $1 \lesssim B \lesssim$ \unit[6]{T} we verified the duality symmetry of our resistance data on either side of critical field $B_c$ following the ``geometric mean" method suggested by Shahar.\cite{SHAHAR} For that we defined several pairs of ``geometrically dual" magnetic fields $B_{sc} < B_c$ and  $B_{ins} > B_c$ such as for each pair  $(B_{sc} B_{ins})^{1/2} = B_c$. Then for each pair we found that $[R(B_{sc})R(B_{ins})]^{1/2} \simeq h/4e^2 $ over a range of temperatures from 140mK to 1K with an accuracy better than 20\%. The details are shown in the Appendix.

Above, in our discussion of $T_0$ in the region of $dR/dT>0$, we ignored the interaction between distant vortices. This allowed us to arrive at a Mott law and interpret the dependence $T_{0} (B)$ observed in experiment. We do not understand why we do not observe the effects of the long range vortex interactions, which would lead to a temperature dependence of resistance that is stronger than Mott's law~\cite{FTY,BOR}. One may be able to resolve this puzzle if, close to critical magnetic field on either side of the SIT, conductivity is due to the hopping of weakly interacting composite fermions made of a Cooper pair and a vortex.\cite{RAGHU}

\begin{acknowledgements}
We are grateful to A. Klein and Han Fu for helpful discussions. This work was supported by the Condensed Matter Physics Program of the National Science Foundation under grant DMR-12663316.  Part of this work was carried out at the University of Minnesota Characterization Facility, a member of the NSF-funded Materials Research Facilities Network via the MRSEC program (DMR-140013), and the Nanofabrication Center which receives partial support from the NSF through the NNIN program.  
\end{acknowledgements}

\appendix
\section*{Appendix: Duality symmetry near the SIT}

Charge-vortex duality is at the root of the dirty boson model of the field-tuned superconductor-insulator transition (SIT).  Duality symmetry describes a case where the transport mechanism within the film is continuous across the transition.  Ovadia \textit{et al.} identify criteria for duality symmetry in the relating $R$ vs $T$ data at a pair of fields $B_{sc}<B_c$ and $B_{ins}>B_c$ which are related to each other and critical field $B_c$ by\cite{SHAHAR}
\begin{equation}
B_{sc}B_{ins}=B_c^{2}. \label{eq:Bpair}
\end{equation}
The symmetry requires the resistances at these fields, $R_{sc}(T)\equiv R_{\square}(B_{sc},T)$ and $R_{ins}(T)\equiv R_{\square}(B_{ins},T)$, to obey
\begin{equation}
R_{sc}=R_{ins}^{-1}. \label{eq:dual}
\end{equation}

\begin{figure}[t]
\centering
\includegraphics[width=0.48\textwidth ]{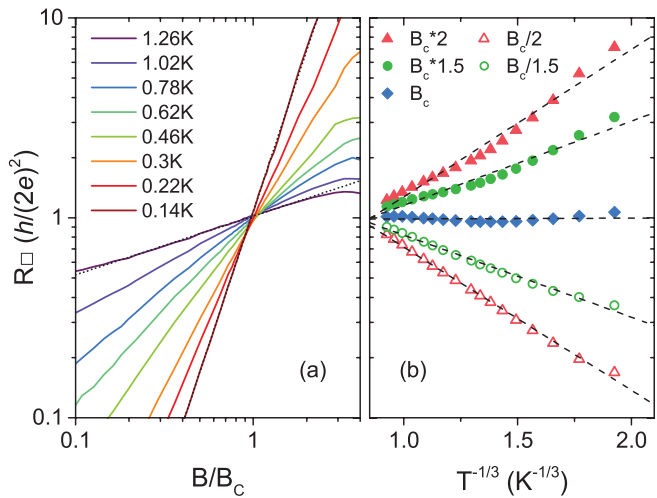}
\caption{(Color online) (a) $R_{\square}$ isotherms as function of perpendicular field, with $B_c=$\unit[2.8$\pm$0.3]{T}.  The dotted lines are fits to the highest and lowest temperature curves to highlight their power law behavior.  (b) $R_{\square}$ vs $T^{-1/3}$ at several paired fields around $B_c$.  The dashed lines show fits to these data using the form for 2D Mott variable range hopping.  Though the fits for fields $B>B_c$ do not agree with this form as well as they do for fields $B<B_c$, they do so well enough to support the observation of duality symmetry across the SIT in this sample.}
\label{Fig:duality}
\end{figure}

Derivation of Eq. \ref{eq:Bpair} in Ref. S6 begins with power-law dependence of resistance isotherms around the crossing point fitting the phenomenological description~\cite{SAMB}
\begin{equation}
R_{\square}(B,T)=R_{c}\left( \frac{B}{B_c} \right) ^{P(T)}, \label{eq:powlaw}
\end{equation}
where $R_{c}$ is the critical sheet resistance of the film, and power $P$ is some function of $T$.  In the case of hopping conduction and this power-law behavior, Eq.~\ref{eq:dual} gives rise to 
\begin{equation}
R_{sc} R_{ins}=R_c^{2}. \label{eq:Rpair}
\end{equation}

The film under discussion does obey power-law behavior at the SIT, as can be seen in Fig. \ref{Fig:duality} (a).  This figure also provides a detailed view of the temperature-dependent crossing points, which are smeared over a range in $B$ and $R$.  This smearing posed a challenge to the determination of $R_c$ and $B_c$.  To calculate these, the set of crossing points were determined for adjacent isotherms from \unit[120]{mK} up through \unit[1.52]{K}. The average of these crossing fields was used as $B_c$ in Eq.~\ref{eq:Bpair} in order to determine sets of paired fields, such as those shown in Fig.~\ref{Fig:duality} (b).  The average resistance at the crossings was similarly used to determine $R_c =$ \unit[6.4]{k$\Omega$} to be used in Eq.~\ref{eq:Rpair} to evaluate the duality symmetry.  Between \unit[140]{mK} and \unit[1.26]{K}, data for fields $B_c/2 \leq B \leq 2 B_c$ were found to satisfy Eq.~\ref{eq:Rpair} well within uncertainty.

\end{document}